\documentclass[twocolumn,showpacs,amsmath,amssymb,prb]{revtex4}
\usepackage{graphicx}
\usepackage{dcolumn}
\usepackage{bm}

\begin{document}

\preprint{APS/123-QED}

\title{Magnetic Interactions and Transport in (Ga,Cr)As}

\author{A. Dakhama}
\affiliation{Department of Physics, Northeastern University, Boston, MA 02115}
\author{B. Lakshmi}
\affiliation{Department of Physics, Northeastern University, Boston, MA 02115}
\author{D. Heiman}
\affiliation{Department of Physics, Northeastern University, Boston, MA 02115}
\email{heiman@neu.edu}
\date{\today}

\begin{abstract}

The magnetic, transport, and structural properties of (Ga,Cr)As are reported.
Zincblende Ga$_{1-x}$Cr$_{x}$As was grown by low-temperature molecular beam epitaxy (MBE).
At low concentrations, x$\sim$0.1, the materials exhibit unusual magnetic properties
associated with the random magnetism of the alloy.  At low temperatures the
magnetization M(B) increases rapidly with increasing field due to the alignment of
\emph{ferromagnetic} units (polarons or clusters) having large dipole moments of order 10-10$^2$$\mu_B$.
A standard model of superparamagnetism is inadequate for describing
both the field and temperature dependence of the magnetization M(B,T).
In order to explain M(B) at low temperatures we employ a \emph{distributed magnetic moment}
(DMM) model in which polarons or clusters of ions have a distribution of moments.
It is also found that the magnetic susceptibility increases for decreasing temperature
but saturates below T=4~K.  The inverse susceptibility follows a linear-T Curie-Weiss law
and extrapolates to a magnetic transition temperature $\theta$=10~K.
In magnetotransport measurements, a room temperature resistivity of
$\rho$=0.1 $\Omega$cm and a hole concentration of $\sim$10$^{20}$~cm$^{-3}$ are found,
indicating that Cr can also act as a acceptor similar to Mn.
The resistivity increases rapidly for decreasing temperature below room temperature,
and becomes strongly insulating at low temperatures.
The conductivity follows exp[~$-$(T$_1$/T)$^{1/2}$~] over a large range of conductivity,
possible evidence of tunneling between polarons or clusters.

\end{abstract}

\maketitle
\section{Introduction}\label{sec:level1}

Utilizing the spin property of electrons is expected to add another
dimension to conventional electronics which relies only on the
charge property of electrons.  The emerging field of \emph{spin
electronics} \cite{Prinz1998, Wolf2001, Awschalom2002} has been
ushered in by the promise of several spin-transport devices.
These include: (\emph{i}) sensitive magnetic field sensors useful for
reading magnetically stored information, based on the
giant magnetoresistance (GMR) effect;\cite{Grunberg1986,Baibich1988}
(\emph{ii}) spin-valves based on the magnetic tunnel junction (MTJ);
\cite{Moodera1995,Miyazaki1995} (\emph{iii}) the spin field effect transistor (spin-FET);
\cite{Datta1990} and (\emph{iv}) magnetic random access memories (MRAM)
utilizing GMR or MTJ.

Even prior to these spin devices, there has been considerable research
aimed at synthesizing new ferromagnetic materials which are compatible
with conventional semiconductors and semiconductor processing. Magnetic semiconductors
have been actively researched for nearly half a century, beginning with
europium chalcoginides (\emph{e.g.} EuX, X=S, Se, Te).\cite{vonMolnar1967,Nagaev1983}
This was followed in the 1980's by II-VI \emph{diluted} magnetic semiconductors
(\emph{e.g.} (Cd,Mn)Te, (Zn,Mn)Se),\cite{Furdyna1988, Dietl1994}
then recently III-V ferromagnetic semiconductors
(\emph{e.g.} (In,Mn)As and (Ga,Mn)As).\cite{Munekata1989,Ohno1998a,Matsukura2002}
Although Ga$_{1-x}$Mn$_x$As possesses robust ferromagnetism
for manganese concentrations near x=0.05, it is only ferromagnetic
below T$_c$$\leq$110.\cite{Ohno1996, DeBoeck1996, Ohno1998a}
More recently there have been reports of higher temperature ferromagnetic semiconductors,
including hexaborides (Ca,La)B$_6$,\cite{Young1999}
phosphides (Cd,Mn,Ge)P$_2$,\cite{Medvedkin2000}
oxides (Ti,Co)O$_2$\cite{Matsumoto2001} and (Zn,V,Co)O,\cite{Ueda2001}
nitrides (Ga,Mn)N,\cite{Reed2001, Sasaki2002}
and antimonides (Ga,Mn)Sb,\cite{Chen2002}.
In addition, using chromium points to high transition temperatures
in III-V materials\cite{Zhao2001, Yamada2002}
and II-VI materials.\cite{Saito2002}  Furthermore, calculations indicate
strong ferromagnetism in (Ga,Cr)As,\cite{vanSchilfgaarde2001}
and the zincblende forms of CrAs,\cite{Akinaga2000} and MnAs.\cite{Sanvito2000}

\smallskip
GaAs doped with Cr was the focus of research a decade ago
because the addition of Cr makes GaAs semi-insulating for use in
electronic applications,\cite{Blakemore1982}
Similar to Fe doping, Cr in GaAs acts as a deep acceptor which compensates
native donors making the material highly resistive.
GaAs:Cr also possesses photoconduction,\cite{Clerjaud1985}
the photorefractive effect,\cite{Imbert1988}
and optically induced change in the Cr valence state\cite{White1980}.
The magnetic properties of (Ga,Cr)As alloys containing substantial concentrations
of Cr are now being explored.  This stems from the ability
to grow GaAs with transition metals
using low temperature molecular beam epitaxy (MBE).\cite{Ohno1996}
The first study of (Ga,Cr)As alloys revealed
superparamagnetic behavior for x=0.03.\cite{Saito2001}
Results have also been reported for (Ga,Cr)As with x=0.11, \cite{Yamada2002}
and CrAs\cite{Mizuguchi2002}.
The present study is aimed at investigating the properties of
Ga$_{1-x}$Cr$_x$As with a Cr concentration of x=0.10.

\smallskip
Magnetic, transport, and structural studies were carried out on samples of
Ga$_{1-x}$Cr$_x$As with x=0.1 grown by low temperature MBE.
Magnetic properties were investigated using a superconducting quantum interference device
(SQUID) magnetometer in magnetic fields up to 5~T and temperatures T=2 to 300~K.
The magnetization M(B) at low temperatures increases much faster for increasing B
than expected for single magnetic ions.  This behavior is evidence of
\emph{ferromagnetic} coupling between magnetic ions.
However, there are many features which cannot be explained by a simple model of
para- or superparamagnetism:
(1) the low field magnetization is nonlinear in B (field dependent susceptibility);
(2) the magnetization deviates strongly from 1/T behavior at low temperatures; and
(3) the magnetization requires a cluster model having a wide distribution
of cluster or polaron magnetic moments.
Although all of the magnetic characteristics cannot be explained by a single model,
some features can be described by a \emph{distributed} magnetic moment (DMM) model
having a large distribution of magnteic moments.

Transport measurements show mild conductivity at room temperature where
$\rho$$\sim$0.1 $\Omega$cm, and strong insulating behavior at low temperatures.
Near room temperature, the conductivity is activated and Hall measurements yield a
hole concentration $\geq$10$^{20}$~cm$^{-3}$.  This indicates that Cr also acts as
a deep acceptor similar to shallower Mn. It is remarkable that the conductivity follows
exp[~$-$(T$_1$/T)$^{1/2}$~] over 8 orders of magnitude change in $\sigma$,
implying a hopping mechanism at lower temperatures.
X-ray diffraction scans exhibit a zincblende structure having a larger lattice constant than GaAs.
\section{Experimental Conditions}\label{sec:level1}

(Ga,Cr)As layers were grown on epiready (100)-oriented
GaAs substrates by low temperature MBE using solid source elements.
Effusion cell temperatures were 980~C for Ga, 275~C for As, and 940~C for Cr.
The Cr-to-Ga flux ratio was monitored by a quartz crystal thickness monitor,
and the As-to-Ga flux ratio was set to approximately 15
by monitoring the flux with a nude ion gauge.  After thermally removing
the surface oxide from the substrate at 630~C for 10-20 minutes in As flux,
a 100~nm thick layer of a high temperature GaAs was grown at 580~C,
followed by a 100~nm layer of low temperature GaAs grown at 220~C,
then the 200~nm thick layer (Ga,Cr)As was deposited at 220~C
at a rate of 0.1~nm/s.
Chromium has higher diffusion than Mn in GaAs,
requiring lower substrate temperatures around 180-220~C
instead of 250~C typically used to grow (Ga,Mn)As.
Cr concentrations were determined by Auger electron spectroscopy
and x-ray photoelectron spectroscopy (XPS).
Magnetization measurements were performed in a variable temperature
5~T superconducting quantum interference device (SQUID) magnetometer.
Plots of the magnetization data were obtained after subtracting the diamagnetism
of the substrate, which was $\chi_{sub}$=-(2.19$\pm 0.01$)x10$^{-7}$~emu/gG.
Four-wire conductivity measurements were made on a standard Hall bar
geometry sample, approximately 4$\times$9~mm in size,
placed in a closed cycle cryostat operating between T=4 and 300~K.
Because of the high sample resistivity at low temperatures,
the DC current was measured while holding the voltage at 1~V.
Hall measurements were made in the cryostat which was placed in a cryogen-free
14~T superconducting magnet having a 52~mm diameter room temperature bore.
\section{Results and Discussion}\label{sec:level1}

\subsection{Field-dependent and Temperature-dependent Magnetization}\label{sec:level2}

\begin{figure}[b] \includegraphics[width=0.4\textwidth]{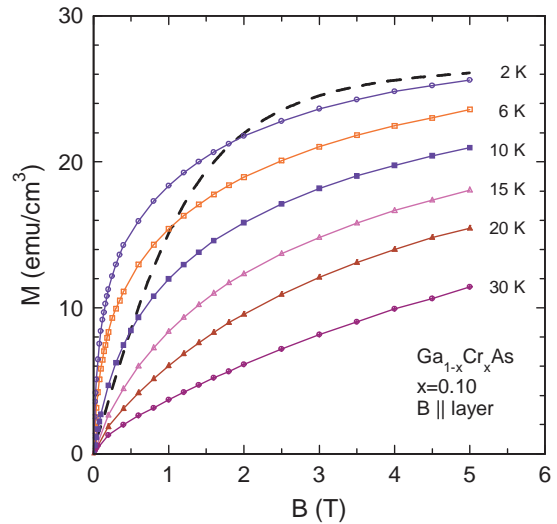}
\caption{Magnetization saturation of Ga$_{1-x}$Cr$_x$As, x=0.095.
The magnetization, M(B), is plotted as a function of applied magnetic field B
for various temperatures from T=2 (top curve) to 30~K (bottom curve).
The field was applied parallel to the epitaxial layer and the substrate diamagnetism
has been subtracted. The dashed curve is a Brillouin function for S=2, g=2, and T=2~K.}
\label{autonum}
\end{figure}

\smallskip

The field dependence of the magnetization, M(B), is shown in Fig.~1
for temperatures ranging from T=2 to 30~K and fields up to B=5~T.
At low temperatures M(B) approaches saturation after several tesla.
The increase of M(B) in Fig.~1 for increasing B is much faster than a paramagnetic response.
The dashed curve is a Brillouin function for T=2~K, representing the response of single (S=2, g=2) ions.
At low fields, the experimental data for T=2~K is many times larger than the Brillouin function.
From this data it is possible to rule out a majority paramagnetism, but it is difficult
to tell whether M(B) is attributed to ferromagnetic or to superparamagnetic response.
The additional magnetism at low fields suggests that the field is aligning \emph{groups}
of magnetic ions rather than single ions.
This behavior is clear evidence of \emph{ferromagnetic} interaction between Cr ion moments.
The interaction couples many ions into ferromagnetic groups of ions having
a large dipole moment which is many times that of a single ion.
Groups of ions in so-called superparamagnets are typically much smaller
in size than domains in ferromagnets and lack their domain wall effects.
The magnetization of individual superparamagnetic blocks have a Langevin function
response, L($\mu$B/kT).
It was found that the present M(B/T) data at various temperatures do not scale with B/T.
This behavior was also observed in a sample with x=0.03 for which M was plotted as a function of B/T,
and the M(B/T) data at various temperatures did not reduce to a single curve.\cite{Saito2001}
We conclude that the behavior of M(B,T) is not that of a simple paramagnet or superparamagnet.

Near B=5~T, the average magnetic dipole moment per ion is found to
be p=$\left\langle \mu  \right\rangle$/$\mu _B$=1.4. This is a
factor of 2 to 3 smaller than expected if all the ions are
aligned. In that case, p=3 or 4, for Cr$^{3+}$(S=3/2, g=2) or
Cr$^{2+}$(S=2, g=2), respectively. The remainder of the
magnetization presumably requires much higher fields to saturate.
The Cr ions giving rise to the missing magnetization probably
exist in a second phase. Although the RHEED and x-ray diffraction
results did not show any appreciable crystal phases other than the
zincblende structure, second phases cannot be ruled out. Another
possibility is that those ions have a different electronic
structure resulting in neither paramagnetic nor ferromagnetic
response to the applied field. Other reported magnetic
measurements of (Ga,Cr)As also find reduced values of p. Values
range from p=2.7 for x=0.009,\cite{Okazawa2001} to p=2.1 for
x=0.034,\cite{Saito2001} and p=1.0 for x=0.11 \cite{Yamada2002}.
This trend of decreasing p for increasing x has been observed for
a range of x up to x=0.065, where it appears that the moment falls
off approximately as p~$\propto$~1/x.\cite{Okazawa2001}

\smallskip
Finally, we note that the present samples do not show hysteresis in the M(B) measurements
for temperatures down to T=1.9~K for either orientation of the magnetic field.
This contrasts with a previous study of an x=0.11 sample which showed
temperature dependent hysteresis.\cite{Yamada2002}  In that study the remanent field
decreased with increasing temperature leading to a transition temperature of T$_c$$\sim$40~K.
Those measurements also contained an unexplained temperature \emph{independent} remanence
which was the same magnitude as the temperature dependent remanence.

\subsection{Modeling of M(B)}

The M(B) dependence has a unique behavior and cannot be fit to any simple function.
Two models are considered here for M(B):
\\\indent (\emph{i}) bound magnetic polaron (BMP) model;
\\\indent (\emph{ii}) \emph{distributed} magnetic moment (DMM) model.\\
In a BMP one itinerant carrier (electron or hole) is bound to a charged center
and there are a number of magnetic ions within the
carrier's orbit.\cite{Wolff1988,Heiman1983}  Because the carrier's orbit size
is predetermined in donors and acceptors, the number of ions (\emph{n}) in each BMP
is about the same for each BMP.  This makes the magnetic dipole moment of all BMPs
equal, except for statistical differences amounting to $\sqrt n$. The \emph{sp-d} exchange
interaction between the carrier and magnetic ions creates a ferromagentic \emph{bubble}.
We propose a DMM model which is similar to the BMP model.
The main difference is that the clusters or polarons can contain more than one carrier, and they
have a broad distribution in size and hence a broad distribution in dipole moment.
In addition, the carriers need not be localized by charged centers fixed
to the lattice {--} the carriers can be localized in groups by Anderson-type disorder.
Also because of the high density of DMMs, they can be magnetically coupled to one another
or even form a percolating network.  Coupled BMPs have been discussed previously
for II-VI diluted magnetic semiconductors.\cite{Durst2002}

\smallskip
Both the BMP and DMM models have a superparamagnetic-like M(B) response, in which \emph{blocks}
of ferromagnetically coupled ion spins with large dipole moments align in the field.
Generally, BMPs and DMMs are \emph{soft} ferromagnets which do not
possess remanence and coercive fields giving rise to hysteresis.
On the other hand, ferromagnetic domains are characteristically different
because of their interesting and important domain wall effects.
Other than that, the magnetism exhibited by these differ primarily
in the scale or size of the discrete magnetic blocks.
In all three cases, the field dependence of the total magnetization usually takes place
in two physically distinct steps:
(\emph{i}) the \emph{total moment} of individual blocks align in a field
even though each block may not be fully saturated at a finite temperature;
(\emph{ii}) the moment within each block increases up to saturation
as the \emph{ions} become fully aligned.
For example, in standard ferromagnets the moment of each domain aligns
in a small field to reach its ''technical'' saturation, followed by
a further increase of magnetization at much higher fields as the moment
of individual domains increases towards full saturation.

\begin{figure}[b] \includegraphics[width=0.35\textwidth]{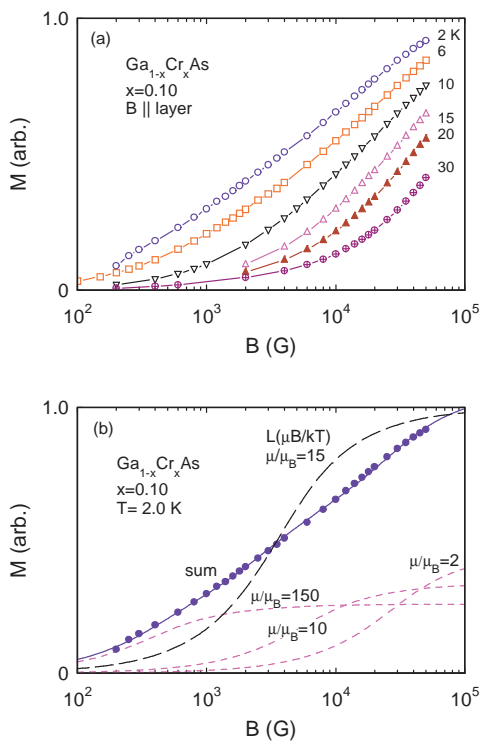}
\caption{Magnetization M versus log(B) of Ga$_{1-x}$Cr$_x$As, x=0.095.
The log of the field is plotted in order to show the low field behavior.
Data for temperatures T=2-30~K are shown in (a).
In (b), M(B) is plotted for T=2~K, where the points are experimental data.
The long-dashed curve is a Langevin function with $\mu$/$\mu_B$=15.
The solid curve shows a fit to the sum of three Langevin functions with magnetic
dipole moments $\mu/\mu_B$=2, 10, and 150.
The three Langevin functions are plotted separately as short-dashed curves.
The field was applied parallel to the epitaxial layer and
the substrate diamagnetism has been subtracted.} \label{autonum}
\end{figure}

\smallskip
Figure~2 shows M(B) for T=2~K, with the field axis plotted on a log scale in order
to display both low and high field behaviors. We first neglect the temperature
dependence of M and focus on the M(B) behavior at T=2~K, displayed in Fig.~2b.
We begin by computing M(B) using a simple BMP model.
In this model the total magnetization is produced by many
equal blocks, each characterized by a large classical magnetic
moment, $\mu$, aligning in a field B at a fixed temperature T.
The magnetization for this, from the Langevin function, is given by
\[
M(B) = M_S \left[ {1/\tanh \left( {\frac{{\mu B}}{{kT}}} \right)
- 1/\left( {\frac{{\mu B}}{{kT}}} \right)} \right],
\]
where $M_S$ is the saturation moment, and $k$=1/11.6~meV/K.
Assuming strong ferromagnetic coupling of $n$ magnetic ions in each block,
their moment is approximated by the sum of the ion moments,
\[\mu=ngS\mu{_B},\]
where \emph{n} is the number of ions which are ferromagnetically coupled,
$g$ the Lande factor, $S$ the spin of individual ions, and $\mu_B$=0.0579~meV/T.
The long-dashed curve in Fig.~2b is a fit to the data points with $\mu/\mu_B$=15.
This value is equivalent to about four S=2(g=2) magnetic ions which are coupled
ferromagnetically.  However, the fit to the data is poor.
There is additional magnetization at low fields, as well as too little at higher fields.
Also, as discussed in the last section, the data for different temperatures
do not scale with the argument of the Langevin function.  Thus, the BMP model
is unable to describe the M(B) data, even at one temperature.
It does establish that some ions are ferromagnetically coupled.

In our DMM model we employ a distribution
in the magnetic moments of the polarons.  The magnetization becomes
\[
M(B) = \sum\limits_\mu  {D_\mu  L_\mu  (B)} ,
\]
where $D_\mu$ is the distribution function of the magnetic moments.
The M(B) data at T=2~K can be fit quite well using only three dipole moments in the sum,
$\mu/\mu_B$=2, 10, 150.  This result is shown by the solid curve in Fig.~2b.
The three separate Langevin functions from the fit are shown by the short-dashed curves.
Although this fit is not unique, it points out that the distribution width
\emph{encompasses several orders of magnitude in magnetic moment}.
This distribution is much broader than the typical $\sqrt n$
ion distribution in fixed diameter acceptor-bound holes in BMPs.
Furthermore, it is also useful to convert the three dipole moments into
corresponding cluster diameters.  For x=0.1 the three dipole moments correspond
to clusters or polarons having diameters of 0.8, 1.3, 3.2~nm, respectively, assuming S=2(g=2) ions.
This factor of 4 in size distribution is not unreasonable considering the sizable
alloy disorder and electrical inhomogeneity in low temperature MBE grown alloys.

\subsection{Temperature-dependent Susceptibility and Ferromagnetism}\label{sec:level2}

\begin{figure}[b] \includegraphics[width=0.37\textwidth]{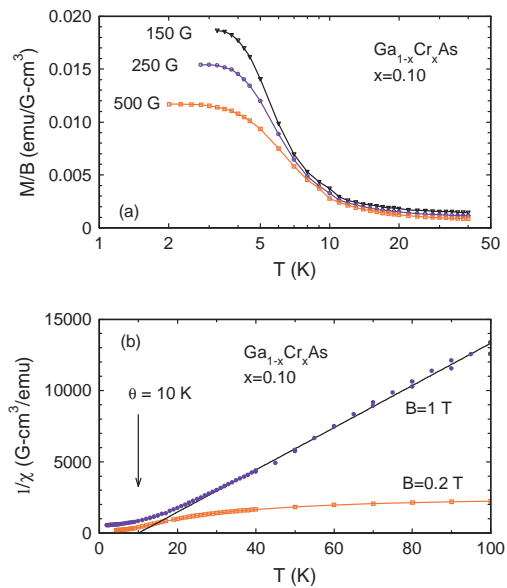}
\caption{In (a), the magnetic susceptibility, M/B, of Ga$_{1-x}$Cr$_x$As,
x=0.095 is plotted as a function of temperature. M(T)/B=$\tilde \chi$(T) was derived from
the magnetization in fields of B=150, 250 and 500~G applied parallel to the
epitaxial layer, and the substrate diamagnetism was carefully subtracted from M.
Below T=4~K, $\tilde \chi$ saturates.
In (b), the inverse susceptibility, $\tilde \chi^{-1}$(T), is plotted for B=1~T.
The straight line represents a fit to the Curie-Weiss law with $\theta$=10.0~K and
Curie constant C=0.00676~emu-K/cm$^{3}$G.}\label{autonum}
\end{figure}

Figure~3a shows the magnetic susceptibility as a function of temperature, $\chi$(T).
The low-field susceptibility was determined from the measured magnetization by
$\tilde \chi$=M(T)/B for fields of B=150, 250, and 500~G. (Note that M/B is not field
independent $-$ the data for the three fields do not coincide at any temperature.)
For all three fields, the susceptibility strongly increases for decreasing temperature.
Below T=4~K, $\chi$(T) flattens out or saturates.
Conventional ferromagnets show saturation in $\chi$(T) when the sample
becomes \emph{demagnetization limited}, and the onset temperature is a lower bound
for long range ferromagnetism.\cite{Shapira1973} The observed saturation of $\chi$(T)
below T=4~K could also be a lower bound for the ferromagnetic transition observed in the
Curie-Weiss behavior discussed below.

\smallskip
The inverse susceptibility was plotted in order to see whether the Curie-Weiss law
describes the paramagnetic response of the magnetic ions at higher temperatures.
Figure~3b shows the temperature dependence of $\chi^{-1}$(T) obtained from M(T) taken at
B=0.2 and 1~T. At temperatures above T=30~K, $\chi^{-1}$(T) is linear in temperature
for the susceptibility measured at B=1~T.
However, $\chi^{-1}$(T) obtained from the low field data is nonlinear in T.
Nonlinearity was also found at lower fields, B=150, 250 and 500~G.  This non-Curie-Weiss
behavior for susceptibility measured at low fields is related to the nonlinearity in M(B)
at low fields, which is due to ferromagnetic response of the magnetic polarons.
The data taken at B=1~T was compared to the Curie-Weiss law,
\[
 \chi  = \frac{C}{{T - \theta }},
\]
where C=$\tilde x$N$_o$p$^2$$\mu _B^2$/3k, p$^2$=g$^2$S(S+1), N$_o$=2.2$\times$10$^{22}$~cm$^{-3}$,
$\mu_B$=9.27$\times$10$^{-21}$~emu/G, and k=1.38$\times$10$^{-16}$~erg/K. The straight line is
a fit to the data for temperatures above T=30~K with $\theta$=10.0~K and C=0.0068~emu-K/cm$^{3}$G.
From the Curie constant, the average dipole moment per ion is p=3.1.  This value obtained
from the paramagnetic behavior is twice that found from the M(B) behavior at low temperatures.
Assuming g=2, the computed average spin is S=1.54, close to S=3/2 for Cr$^{3+}$.
Finally, it is clear from the positive $\theta$ that there are sizable ferromagnetic interactions.

\subsection{Resistivity}\label{sec:level2}

\begin{figure}[b] \includegraphics[width=0.45\textwidth]{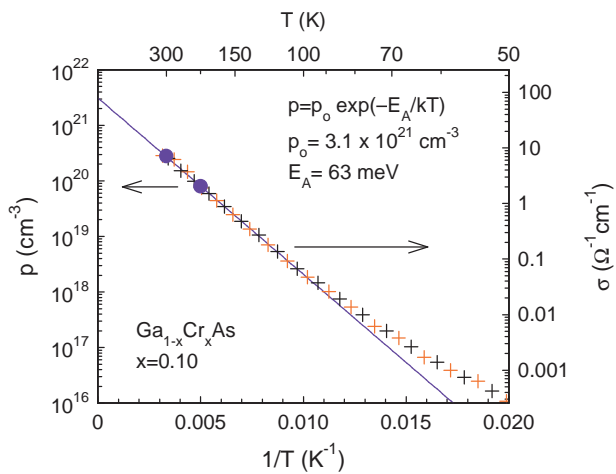}
\caption{Hole concentration and conductivity versus inverse temperature of
Ga$_{1-x}$Cr$_x$As, x=0.095. The solid points are hole concentrations from Hall measurements,
while the crosses are conductivity.  The plotting scales have been shifted to overlap the data.
The straight line is a fit to the data for activated conductivity
$\sigma$=$\sigma _o$exp($-$E$_A$/kT), with an activation energy E$_A$=66~meV.
The conductivity contains data points from both increasing and decreasing temperature sweeps.}
\label{autonum}
\end{figure}

The materials are relatively good conductors at high temperatures, having a small
room temperature resistivity, $\rho$=0.1~$\Omega$cm.
This is in the same range as that observed for (Ga,Mn)As, which has resistivity
about an order of magnitude lower for conducting samples and one to two
orders of magnitude higher in insulating samples.  For decreasing temperature,
the resistivity of (Ga,Cr)As increases by many orders of magnitude and becomes insulating.
This behavior is similar to \emph{insulating} (Ga,Mn)As, which shows insulating
behavior for x$\leq$0.02 and x$\geq$0.06-0.08.\cite{VanEsch1997,Iye1999,Ohno1998a}
Note that (Ga,Mn)As at x=0.02 is both an insulator and shows ferromagnetism.
\cite{Ohno1998a}  The resistivity of (Ga,Cr)As is activated at high temperatures.
In Fig.~4 the log of the conductivity, log($\sigma$), is plotted versus 1/T.
It is clear that log($\sigma$) is linear in 1/T at high temperatures, from T=150 to 300~K.
For this temperature range the activated conductivity follows
\[
 \sigma=\sigma _o~exp(-E_A/kT),
\]
with an activation energy E$_A$=66$\pm$1~meV.  This energy is much smaller than 0.8 eV
for the Cr$^2+$ to Cr$^3+$ acceptor-like transition in GaAs.\cite{Langer1985}
At the high doping levels appropriate to the present samples, the conduction involves activation from
an effective band of electrons formed from the Cr d-levels and disorder-induced band broadening.
Hall measurements were used to estimate the carrier concentration above T=200~K where the conduction
is activated, but Hall measurements are not reliable\cite{Look1990} at lower temperatures where the
conduction is due to hopping.  In the activated region near room temperature, Hall measurements reveal
hole conduction.  The two p(T) data points at T=200 and 300~K have the same slope as $\sigma$(T).
The p(T) data in Fig.~4 extrapolates to p=3$\times$10$^{21}$~cm$^{-3}$, which is close to the density
of Cr ions, xN$_o$=2.1$\times$10$^{21}$~cm$^{-3}$.

\begin{figure}[t] \includegraphics[width=0.40\textwidth]{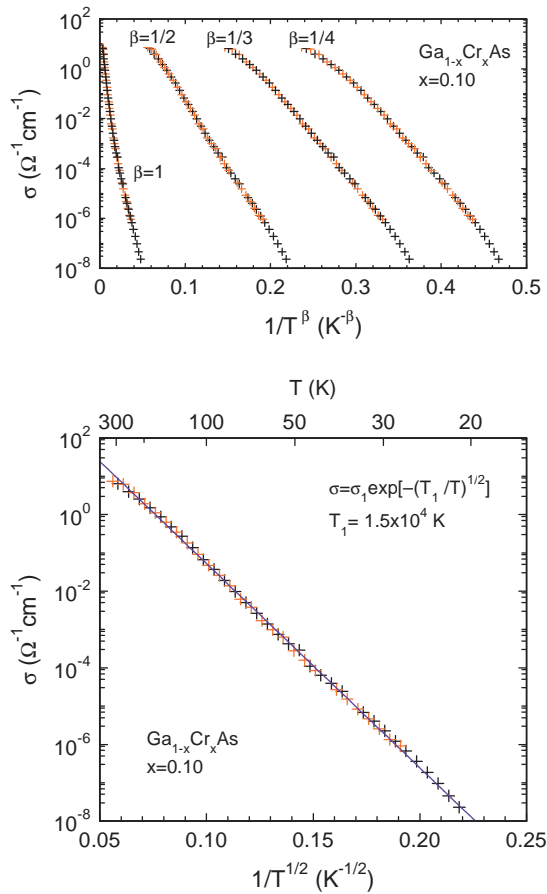}
\caption{Log conductivity versus temperature of Ga$_{1-x}$Cr$_x$As, x=0.095.
The four curves in (a) are plotted on the abscissa as 1/T$^\beta$, where
$\beta$=1, 1/2, 1/3, and 1/4. The best linear relationship corresponds to
$\beta$=1/2 and is shown on an expanded scale in (b).
The straight line fit to the data in (b) represents
$\sigma$=$\sigma _1$exp[~$-$(T$_1$/T)$^{1/2}$~], where
$\sigma _1$=1.1$\times$10$^4$~$\Omega ^{-1}$cm$^{-1}$ and T$_1$=1.5$\times$10$^4$~K.
The plot contains data points from both increasing and decreasing temperature sweeps.}
\label{autonum}
\end{figure}

\smallskip
In order to better understand the conduction mechanisms, $\sigma$ is plotted with
$\beta$=1, 1/2, 1/3, and 1/4 in Fig.~5a.
From these four plots, it appears that the data for $\beta$=1/2 has the highest linearity.
This is displayed on an expanded scale in the lower plot, Fig.~5b.
The data is remarkably linear over the entire range of conductivity of nearly
\emph{nine orders of magnitude}. (Some curvature is seen at higher temperatures where the
conductivity has the exp[1/T] activation.) The straight line in Fig.~5b is a fit to the data using
\[
\sigma (T) = \sigma _1 \exp \left[ { - \left( {T_1/T } \right)^{1/2} } \right],
\]
with $\sigma _1$=1.1x10$^4$~$\Omega^{-1}$cm$^{-1}$ and T$_1$=1.5x10$^4$~K.

A temperature exponent of $\beta$=1/2 has been shown to represent: (\emph{i})variable range
hopping in doped semiconductors having a soft Coulomb gap; or (\emph{ii})tunneling between
conducting regions in granular metals.\cite{Shklovskii1984}  Hopping conduction in (Ga,Cr)As
has been suggested on the basis of the low mobility.\cite{Saito2001}  Also, an exponent of
$\beta$=1/2 has been observed for a sample with x=0.02 over a range of conductivity of 2 and 1/2
orders of magnitude.\cite{Okazawa2001}  In the case of a Coulomb gap,
electron-electron interactions produce a gap in the density of localized states,
with zero density at the Fermi level, E$_F$, and a parabolic dependence on either side of E$_F$.
This model relies on the interaction between localized carriers as they hop from an
occupied state below E$_F$ to an unoccupied state above E$_F$.\cite{Shklovskii1984}
The Coulomb gap mechanism is active when the temperature is less than the gap energy.
The gap energy for Cr could be substantial because of the small radius of the acceptor-like states.
On the other hand, the exponent $\beta$=1/2 can be related to tunneling between conducting regions,
where the conducting objects are the clusters or magnetic polarons which are observed in the magnetization.
Although it is remarkable that $\beta$=1/2 over a range of $\sigma$ of nearly a billion,
the precise mechanism giving rise to the exponent requires further modeling.
However, the mechanism of tunneling between clusters is favored over that of a Coulomb gap
since the magnetization demonstrates the existence of conducting clusters or polarons
which are isolated at low temperatures.
\subsection{X-ray Diffraction}\label{sec:level2}

\begin{figure}[t] \includegraphics[width=0.35\textwidth]{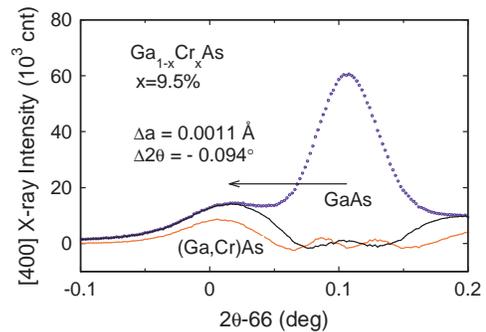}
\caption{X-ray diffraction scan of the (400) reflection of a 200~nm layer of
Ga$_{1-x}$Cr$_x$As, x=0.095, on a GaAs substrate.  The large peak at higher
angle is due to the GaAs substrate and the smaller peak at lower angle is due to the
(Ga,Cr)As layer.  The two solid curves correspond to the data points after subtracting
Gaussian (higher amplitude peak) and Lorentzian functions fit to the GaAs peak.}
\label{autonum} \end{figure}

X-ray diffraction scans did not show any appreciable peaks not related
to the zincblende structure, however, this does not rule out the possibility
of small precipitates or materials having another crystal structure.
The x-ray diffraction spectrum for x=0.10 is shown in Fig.~6.
There is a second peak near 2$\theta$=66$^\circ$in the (400) spectrum.
This weaker peak is down-shifted in angle from the stronger GaAs substrate peak.
The smaller angle corresponds to a larger lattice constant for the (Ga,Cr)As relative to GaAs.
The down-shift of $\Delta2\theta$=$-$0.094$^\circ$ corresponds to a lattice constant
expansion in the growth direction of $\Delta$$a_z$=0.000112~nm.
Similar to (Ga,Mn)As, we expect that layers of (Ga,Cr)As are fully strained
for thicknesses much larger than the critical thickness.\cite{Ohno1998a} In general,
alloys grown at low temperatures have two contributions giving rise to a different lattice constant.
Even without alloying, low temperature growth of GaAs produces a larger lattice constant.\cite{Fatemi1993}
The change in lattice constant of (Ga,Cr)As with Cr concentration has been measured
for x=0 to 0.06, where it was found that d$a_z$/dx=+0.0082~nm.\cite{Okazawa2001}
This expansion is smaller than that for (Ga,Mn)As, where d$a_z$/dx=+0.032~nm.\cite{Ohno1998a}
Note that (Ga,Cr)As has a 4 times closer lattice match to GaAs than (Ga,Mn)As.
However, searching the literature for $a_z$(x) data for (Ga,Mn)As it appear that
$a_z$(x) is not unique and the derivative, d$a_z$/dx, varies by as much as
a factor of two in samples grown in different laboratories.
This means that the x-ray spectrum cannot be used to determine the concentration,
unless perhaps the preparation conditions, such as As/Ga flux ratio, substrate temperature,
growth rate, and post annealing are relatively unchanged.

\section{Conclusions}\label{sec:level1}

(Ga,Cr)As at low Cr concentrations shows anomalous behavior in the magnetic and transport
properties due to the random alloy nature of the magnetic and electronic interactions.
At low temperatures M(B) rises much faster for increasing field than expected for uncoupled
paramagnetic ions.  This is evidence of short-range ferromagnetism between Cr ions.
The M(B) dependence is compatible with a model of local ferromagnetism
in magnetic clusters or polarons having a large distribution in magnetic moment.
However, this model cannot explain the temperature dependence and further modeling is
required to obtain a satisfactory picture of the inhomogeneous magnetism, including
the saturating susceptibility at low temperatures.  A positive $\theta$=10~K from the high
temperature susteptibility is also support for sizable ferromagnetic interactions.
The mechanism for the ferromagnetism is not yet known.  The situation could be similar to the
long range ferromagnetism observed in (Ga,Mn)As, but the smaller hole wavefunction of
the deeper Cr acceptors could give rise to strong localizing effects for quasi-itinerate holes.
It is also possible that double exchange between the deep Cr acceptors could play a major role.
Finally, it is remarkable that the conductivity below room temperature can be described by
exp[~$-$(T$_1$/T)$^{1/2}$~] over a large range of conductivity of 8 orders of magnitude.

\begin{acknowledgments}
We thank C. Feinstein, K. Pant, D.K. Basiaga, G. Favrot. Z. Lee, C. Bailey, T.H. Kim,
and especially J.S. Moodera for considerable help with instrumentation,
and Y. Ohno, S. Kravchenko, R.P. Guertin, G. Berera, and Y. Shapira for useful conversations
and assistance with the measurements. This work was supported by NSF grant DMR-9804313.
\end{acknowledgments}

\bibliographystyle{apsrev}
\bibliography{Dakhama}

\end{document}